\documentclass[iop]{emulateapj}

\usepackage{amsmath}
\usepackage{amssymb}
\usepackage{graphicx}
\usepackage{graphics}
\usepackage{subfigure}
\usepackage{threeparttable}
\usepackage{astron}
\usepackage{natbib}
\usepackage{array}
\usepackage[perpage]{footmisc}
\usepackage{rotating}
\usepackage{textcomp}
\usepackage{color}
\usepackage{listings}
\usepackage{multirow}
\usepackage{wrapfig}
\usepackage{sidecap}
\usepackage{float}
\usepackage{subfloat}

\def\jnl@style{\it}
\def\aaref@jnl#1{{\jnl@style#1}}

\def\aaref@jnl#1{{\jnl@style#1}}

\def\rspsa{\aaref@jnl{Royal~Soc.~of~Lon.~Proceed.~Ser.~A}}                   
\def\aj{\aaref@jnl{AJ}}                   
\def\araa{\aaref@jnl{ARA\&A}}             
\def\actaa{\aaref@jnl{Acta Astron.}}
\def\apj{\aaref@jnl{ApJ}}                 
\def\apjl{\aaref@jnl{ApJ}}                
\def\apjs{\aaref@jnl{ApJS}}               
\def\ao{\aaref@jnl{Appl.~Opt.}}           
\def\apss{\aaref@jnl{Ap\&SS}}             
\def\aap{\aaref@jnl{A\&A}}                
\def\aapr{\aaref@jnl{A\&A~Rev.}}          
\def\aaps{\aaref@jnl{A\&AS}}              
\def\azh{\aaref@jnl{AZh}}                 
\def\baas{\aaref@jnl{BAAS}}               
\def\jrasc{\aaref@jnl{JRASC}}             
\def\memras{\aaref@jnl{MmRAS}}            
\def\mnras{\aaref@jnl{MNRAS}}             
\def\pra{\aaref@jnl{Phys.~Rev.~A}}        
\def\prb{\aaref@jnl{Phys.~Rev.~B}}        
\def\prc{\aaref@jnl{Phys.~Rev.~C}}        
\def\prd{\aaref@jnl{Phys.~Rev.~D}}        
\def\pre{\aaref@jnl{Phys.~Rev.~E}}        
\def\prl{\aaref@jnl{Phys.~Rev.~Lett.}}    
\def\pasp{\aaref@jnl{PASP}}               
\def\pasj{\aaref@jnl{PASJ}}               
\def\qjras{\aaref@jnl{QJRAS}}             
\def\skytel{\aaref@jnl{S\&T}}             
\def\solphys{\aaref@jnl{Sol.~Phys.}}      
\def\sovast{\aaref@jnl{Soviet~Ast.}}      
\def\ssr{\aaref@jnl{Space~Sci.~Rev.}}     
\def\zap{\aaref@jnl{ZAp}}                 
\def\nat{\aaref@jnl{Nature}}              
\def\iaucirc{\aaref@jnl{IAU~Circ.}}       
\def\aplett{\aaref@jnl{Astrophys.~Lett.}} 
\def\apspr{\aaref@jnl{Astrophys.~Space~Phys.~Res.}}
\def\bain{\aaref@jnl{Bull.~Astron.~Inst.~Netherlands}} 
\def\fcp{\aaref@jnl{Fund.~Cosmic~Phys.}}  
\def\gca{\aaref@jnl{Geochim.~Cosmochim.~Acta}}   
\def\grl{\aaref@jnl{Geophys.~Res.~Lett.}} 
\def\jcp{\aaref@jnl{J.~Chem.~Phys.}}      
\def\jgr{\aaref@jnl{J.~Geophys.~Res.}}    
\def\jqsrt{\aaref@jnl{J.~Quant.~Spec.~Radiat.~Transf.}}
\def\memsai{\aaref@jnl{Mem.~Soc.~Astron.~Italiana}}
\def\nphysa{\aaref@jnl{Nucl.~Phys.~A}}   
\def\physrep{\aaref@jnl{Phys.~Rep.}}   
\def\physscr{\aaref@jnl{Phys.~Scr}}   
\def\planss{\aaref@jnl{Planet.~Space~Sci.}}   
\def\procspie{\aaref@jnl{Proc.~SPIE}}   

\begin{document}
\fontsize{9}{9}\selectfont

\bibliographystyle{apj}

\slugcomment{Draft Version 3; \today}

\shorttitle{Resonant Orbits and the High velocity Peaks Towards the Bulge}
\shortauthors{Molloy et al.}

\title{Resonant Orbits and the High velocity Peaks Towards the Bulge}
\author{Matthew Molloy}
\affil{Kavli Institute for Astronomy \& Astrophysics, Peking University, \\ 
Yi He Yuan Lu 5, Hai Dian Qu, Beijing 100871, China (matthewmolloy@gmail.com)}
\author{Martin C. Smith}
\affil{Key Laboratory for Research in Galaxies and Cosmology, Shanghai Astronomical Observatory, \\ 
Chinese Academy of Sciences, 80 Nandan Road, Shanghai 200030, China (msmith@shao.ac.cn)}
\author{N. Wyn Evans}
\affil{Institute of Astronomy, Madingley Road, Cambridge, CB3 0HA, UK (nwe@ast.cam.ac.uk)}
\author{Juntai Shen}
\affil{Key Laboratory for Research in Galaxies and Cosmology, Shanghai Astronomical Observatory, \\ 
Chinese Academy of Sciences, 80 Nandan Road, Shanghai 200030, China (jshen@shao.ac.cn)}

\begin{abstract}
We extract the resonant orbits from an N-body bar that is a good
representation of the Milky Way, using the method recently introduced
by \citet{Molloy2015}. By decomposing the bar into its constituent
orbit families, we show that they are intimately connected to the
boxy-peanut shape of the density. We highlight the imprint due solely
to resonant orbits on the kinematic landscape towards the Galactic
centre. The resonant orbits are shown to have distinct kinematic
features and may be used to explain the cold velocity peak seen in the
APOGEE commissioning data \citep{2012Nidever}. We show that high
velocity peaks are a natural consequence of the motions of stars in
the 2:1 orbit family and that stars on other higher order resonances can contribute to the peaks.  
The locations of the peaks vary with bar angle and, with the tacit assumption that the observed peaks are due to the 2:1 family, we find that the
locations of the high velocity peaks correspond to bar angles in the
range 10$^\circ\lesssim\theta_{\rm bar}\lesssim25^\circ$. 
However, some important questions about the nature of the peaks remain, such as their apparent absence in other surveys of the Bulge and the deviations from symmetry between equivalent fields in the north and south. We show that the absence of a peak in surveys at higher latitudes is likely due to the combination of a less prominent peak and a lower number density of bar supporting orbits at these latitudes.
\end{abstract}

\section{Introduction}

It is now widely accepted that the Milky Way (MW) hosts a bar.  Many
methods have been used to map out the structure of the bar, such as IR
photometry \citep{Blitz1991,Dwek1995}, gas dynamics
\citep{Englmaier1999,Weiner1999}, star counts
\citep{Lopez-Corredoira2007,Robin2012}, microlensing
\citep{Udalski1994,EvBe2002,Wyrzykowski2015} and even the local kinematic landscape
\citep{Dehnen1999,Dehnen2000}.  The MW bar exhibits a boxy-peanut
shape \citep[e.g.,][]{Dwek1995}, which is host to an X-shaped
structure that generates the ``split" red clump \citep{McWilliam2010,Nataf2010,Saito2011,Li2012,Ness2012}.  
Significant observational effort has been expended in mapping the spatial density of the bar using the OGLE-III and VVV data~\citep[e.g.,][]{Wegg2013,Cao2013}. 
More recently, using a large sample of red clump giants, \citet{Wegg2015} derive a bar half-length of between $\sim$4.5 and 5 kpc. However, uncertainties about the nature of the interface between the bar and the disk or spiral arms can strongly influence these estimates \citep{Martinez-Valpuesta2011}. Uncertainty also surrounds some other fundamental parameters of the bar, such as the viewing angle $\theta_{\rm bar}$ and the pattern speed of its rotation $\Omega_{\rm bar}$. The literature reports values for $\theta_{\rm bar}$ in the range $\sim 20^{\circ} \lesssim \theta_{\rm bar} \lesssim 45^{\circ}$ \citep[e.g.,][]{Stanek1997,Benjamin2005} and for $\Omega_{\rm bar}$ in the range $\sim 25 \lesssim \Omega_{\rm bar} \lesssim 50$ km s$^{-1}$ kpc$^{-1}$ \citep[e.g.,][]{Antoja2014,Portail2015a}. 

\begin{figure*}[t]
\begin{center}
\includegraphics[width=\textwidth]{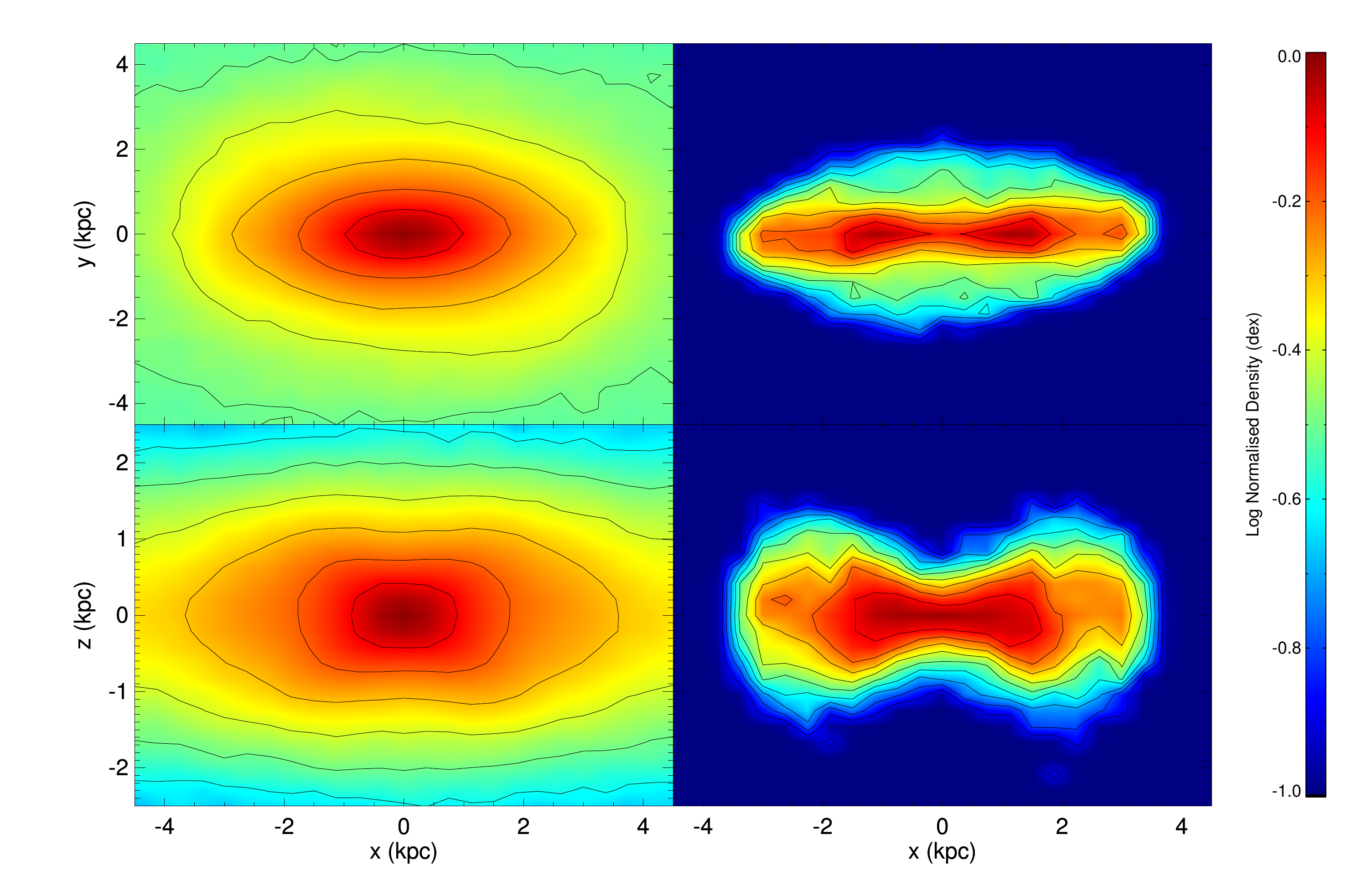}
\caption{The normalised surface density (contours at 0.1 dex) in the $x$-$y$ (top) and $x$-$z$ (bottom)
  planes for all (left) and 2:1 resonant orbits
  (right).}\label{fig:xyDens}
\end{center}
\end{figure*}

Despite a number of
radial velocity surveys toward the Galactic bulge, kinematic
substructure has rarely been observed.  This is a pity, as such
substructure may betray evidence of the processes that formed and
shaped the bar. The Bulge RAdial Velocity Assay 
\citep[BRAVA;][]{Rich2007} and GIRAFFE Inner Bulge Survey
\citep[GIBS;][]{Zoccali2014} both observed $\sim$10,000 giants over a
large region of the bulge, but revealed no signature of cold high velocity peaks.
The ARGOS survey \citep[see][]{Ness2012a,Ness2013b} has also yet to
reveal evidence for streams in the bulge, although their velocity
distributions, cut according to metallicity, hint at the wealth of
information contained in the kinematic data.

Recently, however, a cold high velocity peak has been observed by
\citet{2012Nidever} in the Apache Point Observatory Galactic Evolution
Experiment (APOGEE) commissioning data.  For certain fields towards
the Galactic Bulge, they find bimodal velocity distributions and
identify cold ($\sigma \sim 20$ km s$^{-1}$) secondary peaks in the
distribution of line of sight velocities at $v_{\rm los} \sim$ 200
km s$^{-1}$. The independent observations of \citet{Babusiaux2014} also hint at the presence of a high velocity peak, this time with red clump stars. 
The origin, and even the existence, of this feature has been the
subject of recent debate \citep{Li2014,Zoccali2014}. \citet{Li2014} found the absence of a statistically significant cold high velocity peak in two N-body barred models. They also showed that it is possible for a spurious high velocity peak to appear if there are only a limited number of stars observed.
Here, we look at the matter anew, using novel algorithms to
extract nearly periodic orbits from bar simulations.

\section{Orbital Components in the Bar}

In \citet[][hereafter M15]{Molloy2015}, we introduced a method to
identify resonant orbits in N-body simulations, and used it to provide
a possible explanation for the bimodal velocity distributions observed
towards the Galactic anti-Centre. The method is an alternative to identifying periodic orbits using frequency analysis and is well suited to analysis of evolving N-body models with a changing potential (see M15 for details). It is now applied to
the inner parts of a barred model of the Milky Way. Here, a cold but
thickened disc self-consistently develops a bar, which undergoes a
buckling instability to form a Bulge that is a good match to that of
the Milky Way~\citep[see][]{Shen2010,Li2012,Qin2015}. Resonant orbits can be
characterized by the fact that they close and return to a previously
occupied location in phase space in some rotating frame.  The method
in M15 proceeds by recalculating the N-body orbits in many different
rotating frames.  We define a metric $D_{\rm ps}$ to measure the
distance each particle travels from some arbitrarily chosen starting
point in the rotating frame.  If an orbit almost closes, $D_{\rm ps}$
should, at some point along its trajectory, be nearly zero. By defining a cut-off, we can extract a sample of the nearly closed orbits from the simulation.  Resonant orbits can
librate about each family's parent orbit, so by defining tighter and
tighter cuts on $D_{\rm ps}$, we can extract cleaner and cleaner
samples of resonant orbits.  The choice of cut on $D_{\rm ps}$ is
really set by the problem in hand.

We make some minor modifications to the algorithm described in M15.
In bars, there are always some chaotic orbits, especially near
corotation. Indeed, chaotic orbits may make significant contributions to the structure and dynamics in barred systems. It has been shown that the fraction of chaotic orbits is sensitive to the bar strength and size \citep{Manos2011,Manos2014}. Chaotic orbits may return arbitrarily close to their
chosen starting point over long timescales (the Poincar\'e Recurrence
Theorem), and can therefore be mistaken as periodic.  Previously, we
measured the phase space distance from a single point as the orbit
proceeded on its trajectory over $\sim1$ Gyr. Here, we define a time
frame, unique to each particle, over which we apply the phase space
distance method.  For each particle, we measure the duration it takes
to complete eight radial oscillations. This ensures that we have a
long enough trajectory to extract high-order periodic orbits while
excluding chaotic orbits that rapidly explore their phase space
volume.  For orbits very close to the centre, this time frame may be
sampled by very few points so we interpolate the trajectory. 

\begin{figure}[t]
\begin{center}
\includegraphics[width=0.49\textwidth]{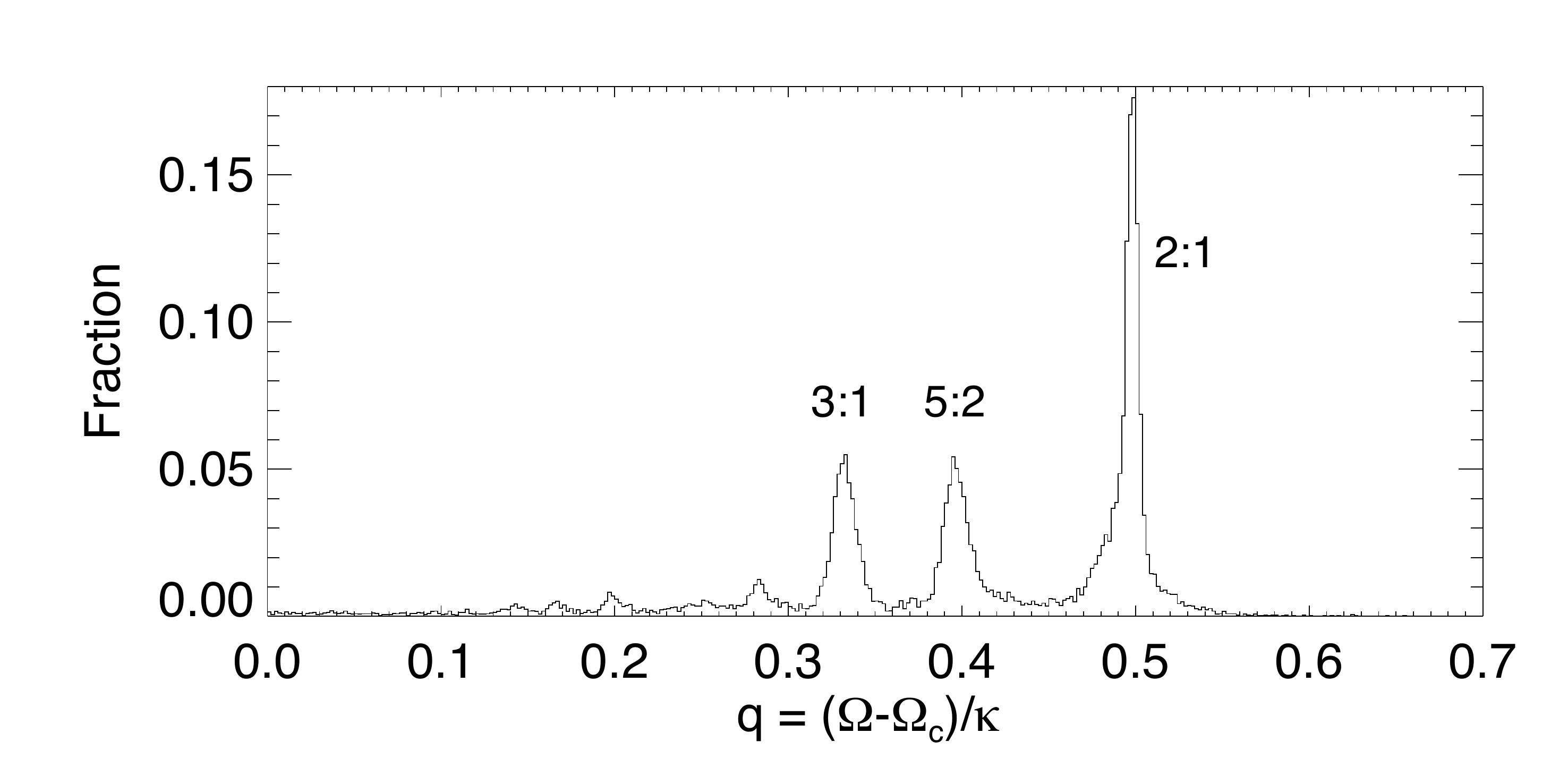}
\caption{The distribution of $q=(\Omega-\Omega_{\rm c})/\kappa$ for the sample of periodic orbits with $R_{\rm g}<2.5$ kpc. We select 2:1 orbits as those with $0.48\le q\le0.52$, 3:1 orbits as those with $0.33\le q\le0.35$ and 5:2 orbits as those with $0.38\le q\le0.42$.}\label{fig:qvalues}
\end{center}
\end{figure}

Also, instead of measuring the phase space distance from a single
point, we find the time at which a particle reaches its first
apocentre, $t_{0}$.  As we scan different rotating frames (between
$37\le\Omega_{\rm p}\le40$ km/s/kpc), we extract sections of the
trajectories that lie in the range $\phi^{\prime}(t_{0})$ to
$\phi^{\prime}(t_{0})+\pi/4$, where $\phi^{\prime}$ is the azimuthal
angle in the rotating frame.  This gives us, say, $n$ similar sections
of the trajectory.  If $n$ is less than three, then we increase the
duration over which we apply the method.  The $n=1$ section is the
reference section which we compare to following test sections.  The
phase space distance $D_{\rm ps}$ is measured between successive
points on the reference section and the test sections.  We take the
average $D_{\rm ps}$ between the orbit sections as a measure of how
``closed" the orbit is - the lower the value, the more closed the
orbit.  Of the $n-1$ $D_{\rm ps}$ values, we take the minimum and in
the following we adopt a cut on the phase space distance of $D_{\rm
  ps}<0.06$ (Over the course of an orbit $D_{\rm ps}$ varies between
$0<D_{\rm ps}<\sqrt{2}$, see M15).

By extracting a sample of resonant orbits in the central regions, we
can deduce the contribution they make to the structure of the bar.
Figure \ref{fig:xyDens} shows the normalized surface density of the
inner parts of the disc.  In the left panels, we show the $x$-$y$
(top) and $x$-$z$ (bottom) surface density for all of the particles in
the simulation.  The bar extends to $\sim$4 kpc or $R_{\rm
  CR}/a\approx1.125$ and has an axis ratio of $b/a\approx0.5$. This is
slightly less extended than the value of 0.35 found by
OGLE~\citep{Rattenbury2007} but agrees well with the structure derived recently from a large sample RR Lyrae \citep{Pietrukowicz2014}. The $x$-$z$ surface density exhibits a strong boxy-peanut shape characteristic of buckled bars.  We extract
the resonant 2:1 orbits by estimating the azimuthal ($\Omega$) and
epicyclic ($\kappa$) frequencies for our sample of closed orbits.  For
each orbit, we calculate $q= (\Omega-\Omega_{\rm p})/\kappa$, where
$\Omega_{\rm p}$ is the pattern speed of the frame in which the orbit
closes, or reaches its lowest $D_{\rm ps}$.  We then extract the 2:1
orbits as those having $0.48\le q\le0.52$.  The surface densities are
shown in the right panels of Figure \ref{fig:xyDens}.  It is clear
that the 2:1 orbits generate the backbone of the bar.  The buckling
instability has a noticeable effect on this family of orbits, inducing
a large vertical velocity dispersion for stars at the end of the bar.
The contribution of the 2:1 family to the boxy and peanut shape is
unmistakable. 

\begin{figure*}[!t]
\begin{center}
\hspace{-1.0cm}\includegraphics[width=0.9\textwidth]{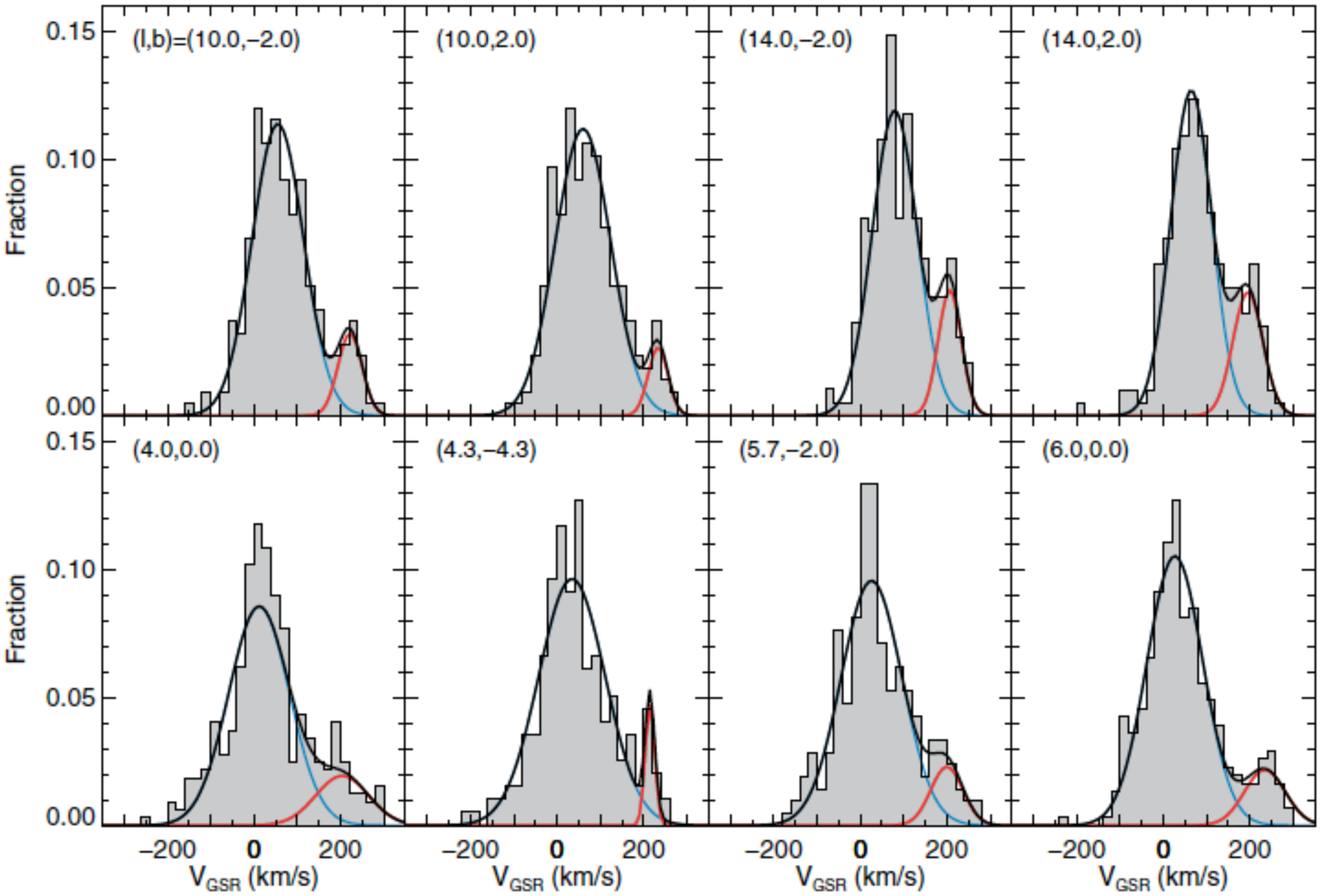}
\includegraphics[width=0.95\textwidth]{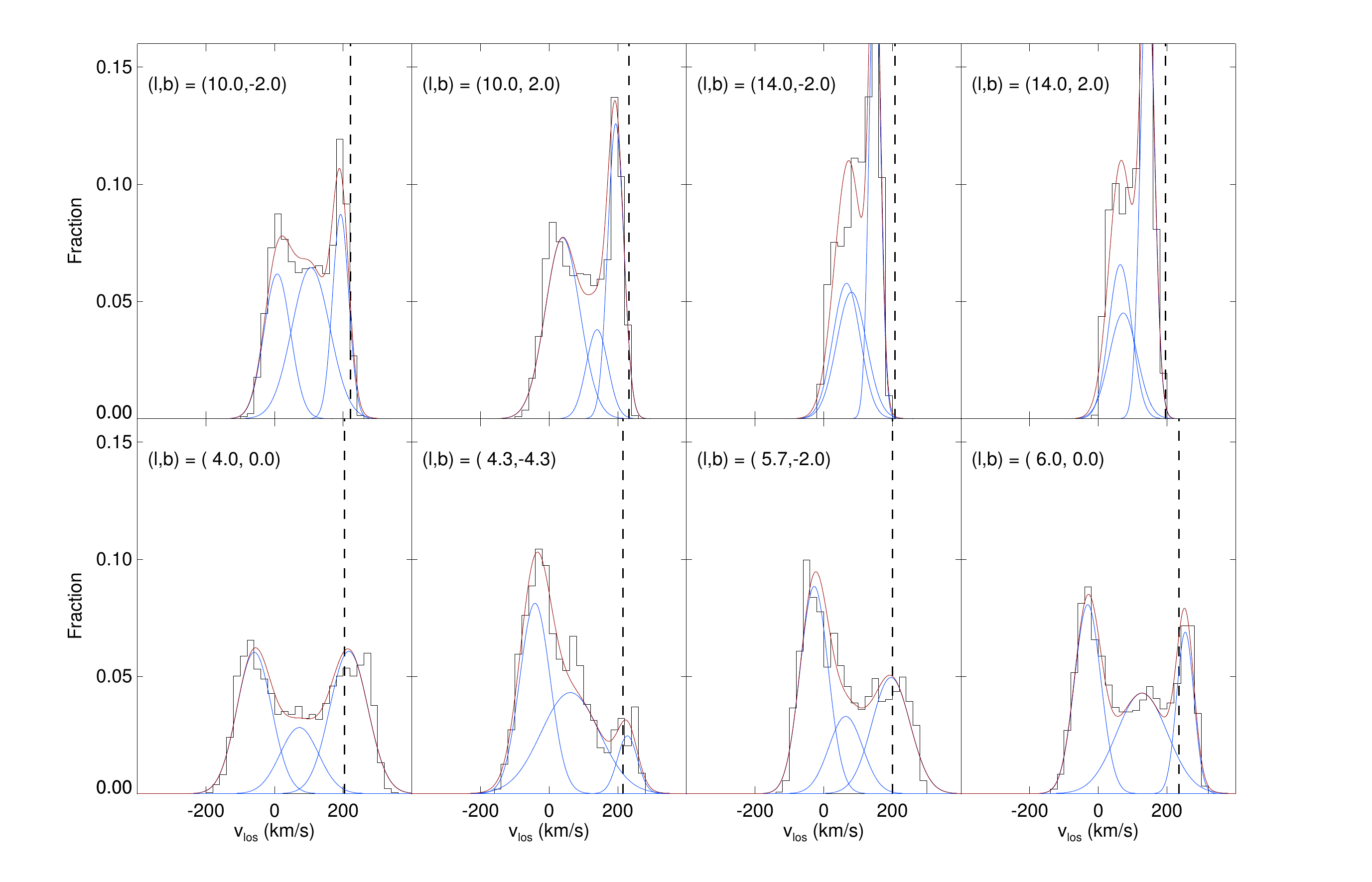}
\caption{\textbf{Top:} For convenience, we show that $v_{\rm los}$ distributions from \citet{2012Nidever}. The blue and red curves show their decomposition of the velocity histogram into two Gaussians. \textbf{Bottom:} The $v_{\rm los}$ distributions for 2:1 resonant orbits in each field, assuming a bar angle of $15^\circ$. A genetic algorithm has been used to populate a 3-Gaussian mixture model for the data. The underlying Gaussians are shown as blue curves and the locations of the peaks reported in \citet{2012Nidever} as the vertical dashed lines. The positions of the peaks show the locations at which 2:1 resonant orbits make the largest contribution to the observed histograms, which of course also include contributions from other orbital families.}\label{fig:NideverFields}
\end{center}
\end{figure*}

\begin{figure*}[!t]
\leavevmode
\begin{center}
\includegraphics[width=\textwidth]{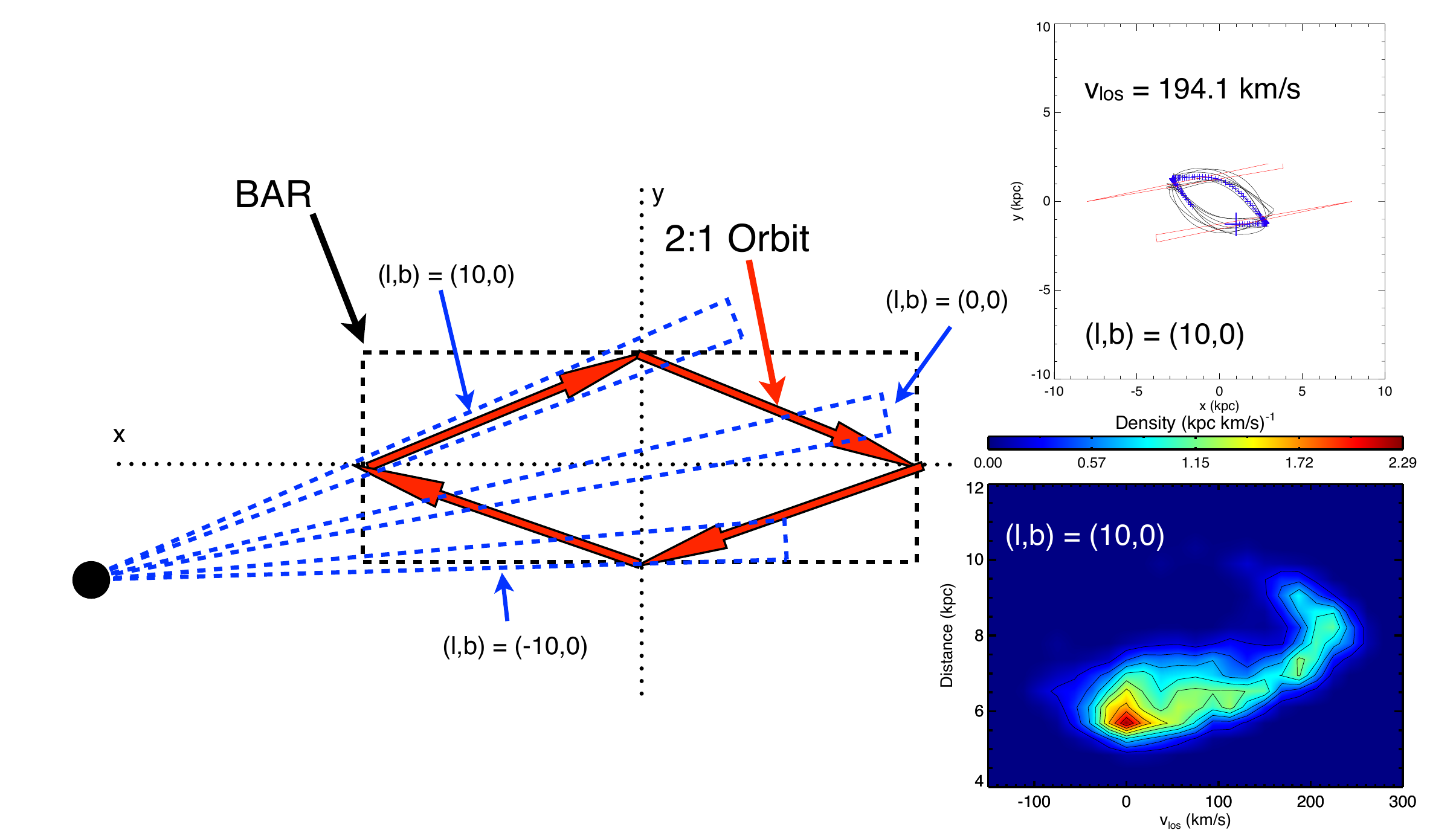}
\caption{A schematic (and simplified) representation of a 2:1 bar
  orbit. The red arrows represent the motion of the particle in the
  frame of the bar (the black dashed box). For different fields,
  represented by the blue dashed lines, it is clear that this type of
  orbit gives different contributions to the line of sight
  velocities.  For $l>0^\circ$, we expect the star to be moving away,
  whilst the reverse is true for fields with
  $l<0^\circ$. \textbf{Inset, top right: }A sample orbit from a
  high velocity peak at $(l,b)=(10,0)$, the red triangles indicate the
  fields of view from equivalent sides of the disc. The blue dots
  represent the preceding 100 timesteps of the orbit and the large
  blue cross, inside the field of view, indicates the position of the
  particle when it has a high $v_{\rm los}$ (listed). \textbf{Inset,
    bottom right: }The density of the 2:1 resonant orbits in the field
  at $(l,b)=(10,0)$ as a function of $v_{\rm los}$ and distance. The
  highest velocity particles lie at a distance of $\sim 8.5$ kpc
  corresponding to pericentric passage. The lowest velocity particles
  lie at a distance of $\sim 6$ kpc corresponding to apocentric
  passage.}\label{fig:21BarOrbit}
\end{center}
\end{figure*}

Of course, other orbit families are present in the simulation. Figure \ref{fig:qvalues} shows the distribution of $q$-values for particles with $R_{\rm g}<2.5$ kpc. The most prominent family are the 2:1 orbits but there is also a significant contribution from the 3:1 ($q\approx0.33$) and 5:2 ($q\approx0.4$) orbit families. The relative fractions for these families is approximately 2:1:1 (where we have taken the 3:1 orbits as those with $0.31<q<0.35$ and the 5:2 orbits as those with $0.38<q<0.42$). The relative fractions between the orbit families remains largely unchanged using different cuts on $D_{\rm ps}$. However, the normalisation of the phase space distance (see M15) affects the elongated 2:1 orbit family differently compared to the more circular 3:1 and 5:2 orbits. Consequently, we refrain from making any strong conclusions based on the relative fractions of the extracted periodic orbits. In any case, we choose a cut on the phase space distance that allows us to sample the phase space of each family sufficiently. So, while the relative fractions of each family is uncertain, the shapes of the distribution functions are robust. 

\begin{table}
  	 \centering 
 \begin{threeparttable}
  \caption{Line-of-sight velocity distributions for each of the APOGEE commissioning fields (truncated). Table 1 is published in its entirety in the electronic edition of ApJ, a portion is shown here for guidance regarding its form and content.}
   \begin{tabular}{c | c | c | c }
    \hline \hline

 (l,b) & $\theta_{\rm bar}$ & All Peak\tnote{a} & 2:1 Peak\tnote{b} \\ 
 ($^\circ$) & & km s$^{-1}$ & km s$^{-1}$  \\ \hline 
 
 \multirow{4}{*}{(10.0,-2.0)}  & 10$^\circ$  & 60.2  & 192.9 \\
   & 15$^\circ$  & 55.2  & 192.4 \\
   & 20$^\circ$  & 56.3  & 175.6 \\
   & 25$^\circ$  & 51.9  & 180.9 \\ \hline
 
  \multirow{4}{*}{(10.0,2.0)} & 10$^\circ$  & 70.1  & 197.3 \\
   & 15$^\circ$  & 64.7  & 193.9 \\
   & 20$^\circ$  & 62.5  & 197.7 \\
   & 25$^\circ$  & 56.3  & 183.9 \\ \hline \hline
 
%
%
%
%
%
 
\hline
\end{tabular}\label{tab:Peaks}
 \begin{tablenotes}
	\item[a] $\chi^{2}$ fits to all particles in the field.
	\item[b] Maximum peak of 3-Gaussian fit. 
  \end{tablenotes}
  	 \centering 
  \end{threeparttable}
\end{table}

\begin{figure*}[!t]
\begin{center}
\includegraphics[width=0.85\textwidth]{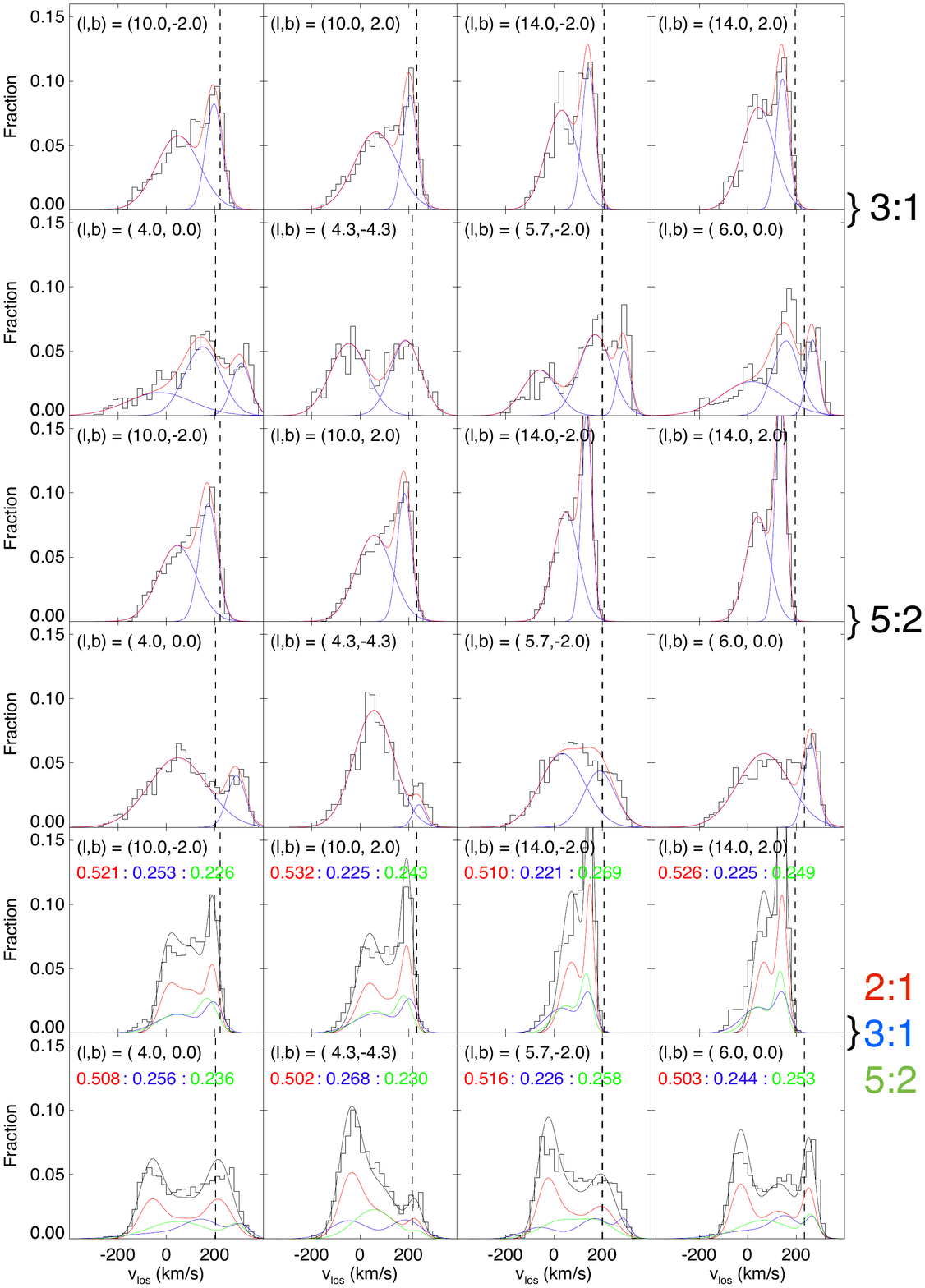}
\caption{The $v_{\rm los}$ distributions for the 3:1 (top two rows) and 5:2 (middle two rows) orbit families in the APOGEE commissioning fields. For the 3:1 and 5:2 families we average over 35 \& 50 timesteps ($\sim$0.34 \& 0.48 Gyr) respectively to generate the distributions, which are then modeled with GMMs. The location of the observed peak for each field is indicated by the vertical dashed line \citep{2012Nidever}. The distributions due the 2:1, 3:1 and 5:2 families combined are shown in the bottom two rows (where we have averaged over 10 timesteps or $\sim$100 Myr). The relative fractions of each family are listed inset (2:1--red, 3:1--blue, 5:2--green). We overplot our simple model which uses relative fractions of 2:1:1 (for the 2:1, 3:1 and 5:2 families respectively) as the black curve. The components, using these relative fractions, are shown as the coloured curves where we use the 3 Gaussian fit to the 2:1 family from Figure \ref{fig:NideverFields} and the fits to the 3:1 and 5:2 families from the top four panels.}\label{fig:multivlos}
\end{center}
\end{figure*}

\section{Velocity Distributions}

The photometry and star counts for this model have already been shown
to be a good match to observations.  Indeed, the simulation was
tailored to match the kinematics towards the Galactic Bulge as seen by
the BRAVA data~\citep{Shen2010}.  
The pattern speed of the simulation bar stays roughly constant at $\sim$40 km s$^{-1}$ kpc$^{-1}$ (see M15). This is in the middle of the range of values reported in the literature and is consistent with the most recent estimates from gas dynamics \citep{Sormani2015}\footnote{\citet{Aumer2015} have recently shown that the $v_{\rm los}$ distributions are only weakly affected by the pattern speed, at least between 25 and 30 km s$^{-1}$ kpc$^{-1}$.}. A suite of simulations spanning the parameter space of bar size, strength, viewing angle and pattern speed is required to fully investigate their effect on the $v_{\rm los}$ distributions. 

Since we can deconstruct the bar
into its different orbital families, we can now characterize the
contribution of each family to the velocity distributions.  The APOGEE
commissioning data revealed cold high velocity peaks for a number of
Bulge fields. \citet{Binney1991} were the first to link the motions of gas towards the Bulge to orbits in (planar) elliptical potentials. Below, we will show that these peaks arise
naturally as a result of the motions of resonant bar orbits, in
particular, the 2:1 orbital family. Note that if we include all simulation particles in our $v_{\rm los}$ distributions we recover the results of \citet{Li2014}, in which no cold peaks are found. 

To generate mock $v_{\rm los}$ distributions, we first fix some
fundamental parameters.  We choose the Solar radius as $R_{0}=8.5$
kpc and the circular velocity at $R_{0}$ as $v_{\rm c}=220$
km s$^{-1}$. Varying these between reasonable values has only minor
effects on the distributions.  We assume an angle between the long
axis of the bar and the Solar--Galactic Centre (GC) line of
$\theta_{\rm bar}=15^\circ$ (we later justify the choice, where we explore a range of bar angles).  We also limit the distances of the particles to between 3 kpc and 9 kpc
and, in order to increase the numbers in the samples, we include
particles from equivalent positions on either side of the disc and
increase the diameter of the field by a factor of two compared to the
APOGEE fields. As a further measure to increase the number of
particles, we also average over 10 timesteps, making sure to take into
account the (small) change in bar angle between timesteps.

The top panel of Figure~\ref{fig:NideverFields} shows the kinematic
data on the APOGEE commissioning fields of \citet{2012Nidever},
together with their two Gaussian decomposition.  In the bottom plot of
Figure \ref{fig:NideverFields}, we show the velocity distributions for
our sample of 2:1 orbits.  To avoid forcing fits to binned data, we
instead opt for a more general approach.  We populate Gaussian mixture
models (GMMs) using a genetic algorithm that converges on probability
distribution functions that could have produced the data\footnote{We
  do this using the freely available SOLBER routines:
  http://www.ast.cam.ac.uk/$\sim$vasily/solber/}.  This
only requires one to input the range of parameter space to explore.
In order to compare the fits across each field, we force the
distributions to be fit with three Gaussians. Initially, we fit each
of the field's distributions with one to five Gaussians. We then
performed likelihood ratio tests to see when adding an extra Gaussian
component made no significant improvement. Most of the fields
preferred either two (43.75\%) or three (52.5\%) Gaussians, while only
a small proportion preferred one Gaussian (3.75\%). Generally, the distributions are split into negative and positive velocity components, with an intermediate component in some fields. We interpret the negative velocity component as being due to particles on the near side of the bar, streaming towards apocentre. The high velocity component are then the particles streaming towards pericentre on the far side of the bar, while the intermediate component represents the particles that are slowing down as they approach apocentre, those at apocentre (with almost zero line-of-sight velocity) and those leaving apocentre, speeding up as they head towards pericentre. The shape, and number of components in the distribution is a non-trivial function of the field being observed, the chosen bar angle and the range of distances being sampled. For a selection
of bar angles, we list the values of the peaks of the distributions
for each field in Table \ref{tab:Peaks}. In the case of the three
Gaussian fits to the 2:1 distributions, we list the highest valued
peaks.

A crude representation of a 2:1 bar orbit is shown in
Figure \ref{fig:21BarOrbit}. The orbit, shown as the red arrows,
reaches its apocentre at each end of the bar (on the $x$-axis), while
the pericentres are located in the directions perpendicular to the bar
(the black dashed box).  The black dot represents the position of the
Sun, so that the bar is rotating in a clockwise direction and the
Galactic Centre-Solar position line is at an angle close to $20^\circ$ with respect to the bar.  The dashed blue lines
indicate the fields of view in Galactic coordinates.  With
$(l,b)=(10^\circ,0^\circ)$, we can see how this type of orbit contributes to a
high velocity component in this direction.  For this line-of-sight,
the measured $v_{\rm los}$ captures most of the components of
Galactocentric $v_{\rm R}$ and $v_{\phi}$ and results in the high
velocity peaks shown in Figure \ref{fig:NideverFields}.  In the
Galactocentric frame, these stars have $v_{\rm R}<0$ km s$^{-1}$ and
$v_{\phi}\ll0$ km s$^{-1}$ which, for this particular direction in relation
to the Solar position, make a significant contribution to $v_{\rm los}$ -- in fact, almost all of the particle's velocity is coincident
with the line of sight.

\begin{figure*}[t]
\begin{center}
\leavevmode
\includegraphics[width=\textwidth]{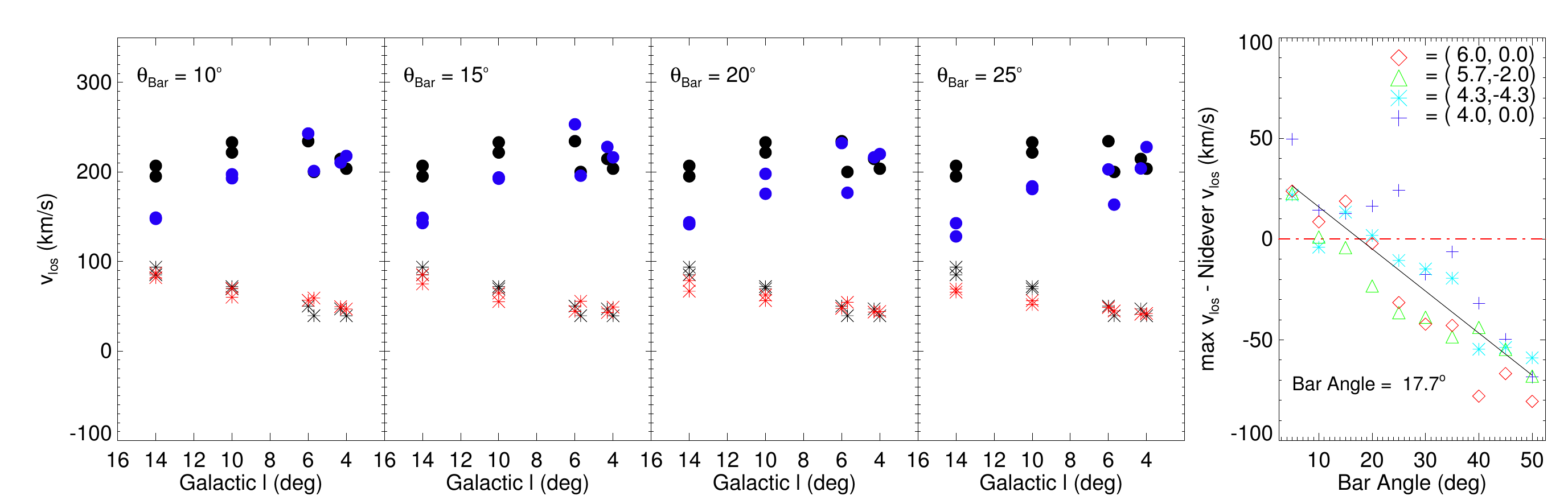}
\caption{\textbf{Left Panels: }For the APOGEE commissioning fields shown in
  \citet{2012Nidever}, we use Gaussian mixture models to characterize
  the $v_{\rm los}$ distributions assuming bar angles in the range
  $5^\circ\le\theta_{\rm bar}\le50^\circ$ (shown here is
  $10^\circ\le\theta_{\rm bar}\le25^\circ$). The black
  stars/dots represent the general/high velocity peaks seen in the
  APOGEE data (note that some longitudes have two points since APOGEE observed fields symmetric about the plane). The blue dots are the highest velocity peaks from our
  Gaussian mixture models for the 2:1 orbit velocity
  distributions. The red stars are the peaks from $\chi^{2}$ fits to
  all particles in the fields. \textbf{Right Panel: }The difference in the values of the simulation and observed peaks for fields with $l<10^{\circ}$ as a function of $\theta_{\rm bar}$, to which a linear fit is made (black line). The intersection of the fit to the 0 km/s difference suggests a bar angle of $\theta_{\rm bar}=17.7^{\circ}$.}\label{fig:Pos_Vel}
\end{center}
\end{figure*}

As a specific example, we take a sample orbit in the direction of
$(l,b)=(10^\circ,0^\circ)$ from the high velocity component ($v_{\rm los}>150$
km s$^{-1}$) in the velocity distributions assuming a bar angle of
$20^\circ$.  We plot the orbit in the upper right panel of Figure
\ref{fig:21BarOrbit} and indicate with the red lines the fields of
view from which our sample is derived.  The orbits are plotted over a
period of $\sim$1 Gyr with the final 100 timesteps indicated with blue
crosses and the final timestep shown as the large blue cross lying
inside the field of view.  The line of sight velocity, assuming a
Solar position of $(x,y)=(8.5,0)$, is also listed.

A population of near-circular orbits would make a similar, but smaller, 
contribution to the high velocity peak in $v_{\rm los}$ since $v_{\rm R}\approx0$ km s$^{-1}$. This direction also captures stars that are just reaching their maximum radius and so contribute to the negative velocity component in the distributions of
$v_{\rm los}$.  The peaks of this component are at a lower velocity
compared to the simple disc rotation model.  This shows that resonant
orbits in this direction imprint both low and high velocity kinematic
signatures on the line of sight velocity distributions.  A field
centered on $(l,b)=(0^\circ,0^\circ)$ passes through the whole
structure of the bar and therefore catches stars with a negative
$v_{\rm los}$ on the near side and with positive $v_{\rm los}$ on the
far side.  At negative $l$, the bar stars are approaching the Solar
position and so imprint a high negative velocity component, mirroring
the corresponding fields at positive $l$.  Having a non-zero bar angle
influences the differing shapes of the distributions between positive
and negative longitude fields.

The lower right panel of Figure \ref{fig:21BarOrbit} shows the density
of particles as a function of $v_{\rm los}$ and distance.  For the
line of sight ($l, b$) = ($10^\circ, 0^\circ$), particles that are
reaching apocentre contribute to the peak at $\sim$ 0 km s$^{-1}$ and
are at smaller distances ($\sim$6 kpc).  As the distance is increased,
the particle's velocity increases since both $v_{\phi}$ and $v_{\rm
  R}$ are increasing and also because the velocity vector is
coincident with the line of sight. Of course this is only true for the 2:1 orbits, the wide range of other orbit families in static and evolving potentials will have more complicated morphologies \citep[e.g.,][]{Pfenniger1991,Manos2014}. The highest velocity particles are
approaching their pericentre occurring on the short axis of the bar at
a distance of $\sim$8 kpc, very close to $R_{0}$.  This picture is
consistent with our interpretation of the high velocity peaks in Bulge
fields being due to the motions of resonant 2:1 bar orbits. 

The other orbit families are also likely to have an influence on the $v_{\rm los}$ distributions. We see from Figure \ref{fig:qvalues} that the 3:1 and 5:1 families make up a significant portion of our periodic orbits. If we plot the $v_{\rm los}$ distributions for these orbits it is clear that they are rich in structure (Figure \ref{fig:multivlos}). To account for their lower numbers, we average over 35 and 50 timesteps (336 Myr and 480 Myr) for the 3:1 and 5:2 families respectively. In this way, each of the chosen particles moves around their respective orbit pattern roughly twice. For fields with $l\ge10^{\circ}$ the 3:1 (top two rows) and 5:2 (middle two rows) families also generate strong high velocity peaks (as before, we use a likelihood calculation to choose the number of Gaussian components). Just as for the 2:1 families, the peaks in these fields lie at a lower velocity compared to the peaks identified by \citet{2012Nidever} (vertical dashed lines). At lower longitudes ($l\le6^{\circ}$) these families also generate peaks coincident with the peaks identified in the data (e.g., 3:1 and 5:2 at $(l,b)=(5.7^{\circ},-2.0^{\circ})$). However, there are fields in which the peaks fail to match the data (e.g., 3:1 and 5:2 at $(l,b)=(4.0^{\circ},0.0^{\circ})$). For certain bar angles $\theta_{\rm bar}$ the 2:1 family can simultaneously, across a number of fields, generate peaks that are a good match to the data. 

The relative fractions of each orbit family is an important factor in shaping the $v_{\rm los}$ distributions (i.e., the locations of cold peaks). It is likely the case that the relative fraction in each field differs from the ``global" fraction for the bulge as a whole. This is because each family has a distinct spatial density distribution. For example, the 2:1 family stays close to the long axis of the bar while the other families generally travel farther along the short axis. This, together with the uncertain distances probed in the APOGEE fields, makes it very difficult to derive the relative fractions from the data. We show the $v_{\rm los}$ distributions imprinted by the 2:1, 3:1 and 5:2 families combined in the bottom two rows of Figure \ref{fig:multivlos}. The combined 2:1, 3:1 and 5:2 $v_{\rm los}$ distributions average over 10 timesteps. As a fiducial model we apply simple relative fractions to the fits for each family (2:1:1 for the 2:1, 3:1 and 5:2 families respectively; for the 2:1 family we use the 3-Gaussian fits from the previous section) and overplot the simple model on the $v_{\rm los}$ distributions. The actual fractions from the model, listed in each panel, vary only slightly from field to field but for this exercise we have applied a uniform distance cut along each field. The distances probed in the data are likely to vary significantly with Galactic $l$ and $b$. Given the small change in fractions between fields, it's not surprising that the simple model matches the distributions quite well. With this model the high velocity peak is largely determined by the 2:1 family, especially for the fields with $b=0^{\circ}$. At larger latitudes, the other families appear to make more significant contributions to the high velocity component, but never stronger than the 2:1 family. 

With the tacit assumption that the locations of the peaks are set by the 2:1 family (i.e., that the relative fractions of the most populated families is close to 2:1:1, as above), can we use this insight to constrain the viewing angle of the bar,
using the information on how the velocity peaks vary with Galactic
position? In Figure~\ref{fig:Pos_Vel} (left panels), the black stars and dots
represent the locations of the peaks found by \citet{2012Nidever}.
Over-plotted as blue dots and red stars are the positions of the peaks
from our models. Specifically, the blue dots are derived from the
Gaussian mixture models for the 2:1 resonant orbits, whilst the red
stars are extracted from $\chi^{2}$ fits to all the particles in the
field. It is clear that, although no single choice of viewing angle
reproduces all the data, the trends in the velocity peaks are
well-reproduced for bar angles $\sim 15^\circ$. 

\begin{figure*}[t]
\begin{center}
\leavevmode
\includegraphics[width=\textwidth]{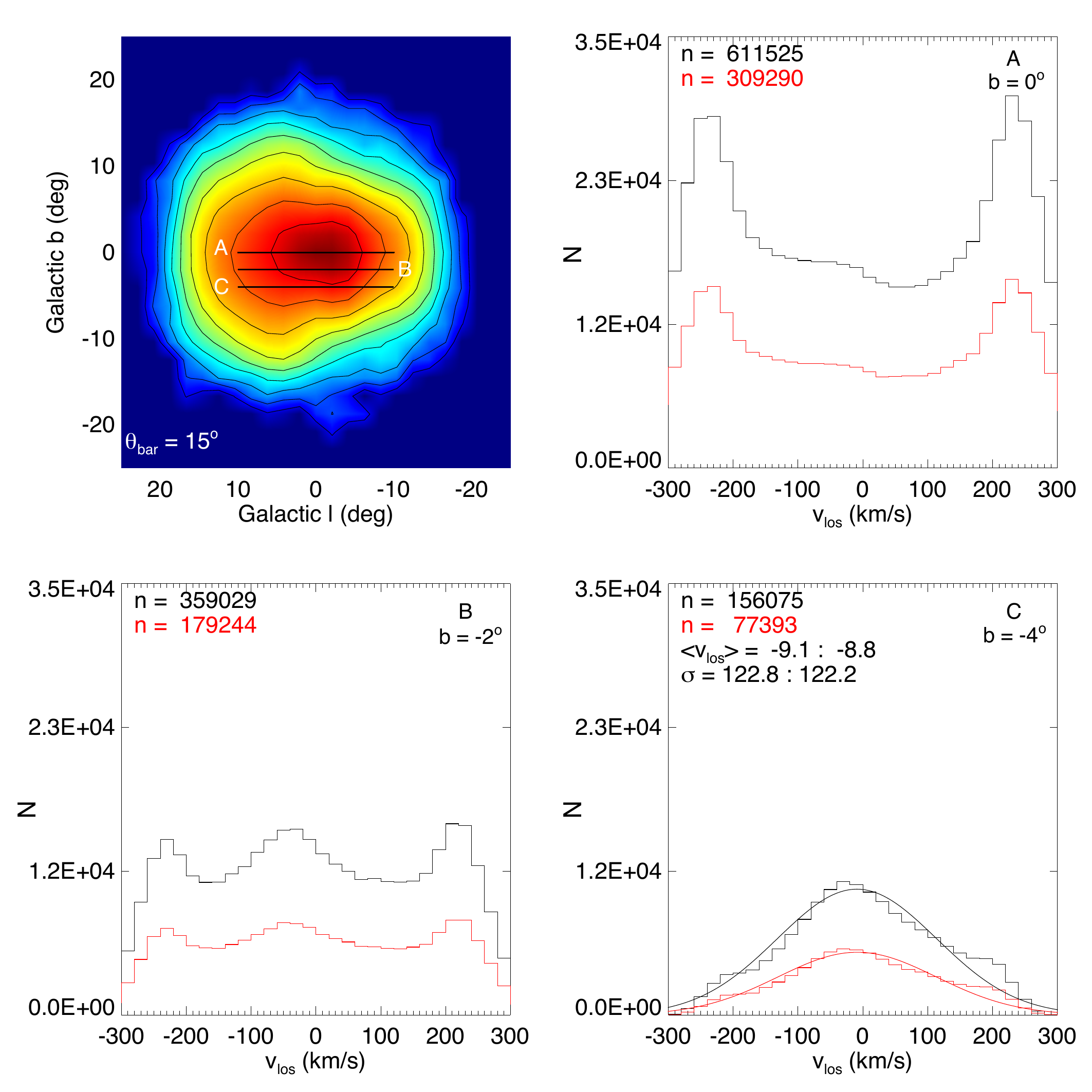}
\caption{\textbf{Top Left: }The projected surface density of the 2:1 resonant orbits assuming a bar angle of $\theta_{\rm bar}=15^{\circ}$ (averaged over 0.48 Gyr). Horizontal lines indicate stripes in longitude from which the following histograms are drawn. \textbf{(A)} $v_{\rm los}$ distribution for the strip with $-10^{\circ}<l<10^{\circ}$ and $b=0^{\circ}$. The black histogram represents a strip of width $\triangle b=1^\circ$ (thick) and the red, a strip of width $\triangle b=0.5^\circ$ (thin). \textbf{(B)} As for panel A, but at $b=-2^{\circ}$. \textbf{(C)} As for panel A, but at $b=-4^{\circ}$. We also make $\chi^{2}$ fits to the distributions, the mean and dispersion of which are listed for the thick and thin strips respectively. }\label{fig:vlos_b}
\end{center}
\end{figure*}

The fields with the largest deviations are the ones with the highest longitude, for which our simple picture probably breaks down. 
Although the peaks in these fields are quite pronounced, they systematically lie at lower values compared to the data. 
This could be due to one of two possible scenarios. 
Firstly, being at high longitudes, particles in these fields feel a significant cumulative effect of the shallower potential. 
The high velocity particles in fields with $l\ge10^{\circ}$ are at pericentre between 1.5 and 2.0 kpc along the short axis of the bar. 
We expect the potential to be somewhat shallower since the pure disc simulation is absent of a gaseous component and live halo that may relax into a more concentrated configuration after the formation of the bar. 
However, the simulation has been shown to be in good agreement with the kinematics observed by BRAVA \citep{Shen2010}, even as far out as $l=10^\circ$. 
A good match is made to the mean velocities and velocity dispersions, so that the comparison is made to the data through the whole line of sight. 
As we've shown above (Figure \ref{fig:21BarOrbit}, bottom right), the high velocity peaks correspond to a limited range in distance. 
That the deviation from the observed peaks increases with $l$ is another indication that a somewhat shallow potential is the cause.  

Another possible scenario is that the peaks are in fact caused by another family of resonant orbits. 
We have checked the other major families in the bar, the 3:1 and 5:2 families. 
They do generate strong peaks in these fields but, as with the 2:1 orbits, the high velocity peaks are at systematically lower values (combinations of the 2:1, 3:1 and 5:2 orbits also result in peaks with low values, see Figure \ref{fig:multivlos}).  
However, other higher order resonances may also be important in these regions. 
As mentioned in \citet{2012Nidever}, the high velocity peaks are unlikely to be due to tidal streams. 
Although the Sagittarius stream lies close by on the plane of the sky, the high velocity stars show no preference for magnitude or metallicity and, in any case, the stream stars are not expected to appear in large numbers \citep{Law2010}. 

Another way of synthesizing
this information is presented in the right panel of Figure~\ref{fig:Pos_Vel}, which
shows how the difference between the observed and simulation high velocity peak varies with assumed viewing angle of the bar. The observational data for each field are
represented by the horizontal red dot-dashed line (i.e., the zero-difference line), and suggest viewing angles between $10^\circ$ and $25^\circ$. Some scatter is expected, as the N-body model does not exactly reproduce the three-dimensional density
of the inner Galaxy and the kinematical properties are subject to
numerical shot noise. 

\section{Discussion}

We have shown that the high velocity peaks seen in the APOGEE
commissioning data may be explained by the
presence of a large family of 2:1 resonant orbits in the Galactic
bar, as was tentatively suggested by \citet{2012Nidever}. These orbits are elongated along the bar's major axis, and must
support the distended shape of the bar to provide its backbone. Indeed, it has already been shown by \citet{Binney1991} that the motion of gas towards the Bulge follows naturally from orbits in (planar) ellipsoidal potentials. When
viewed at bar angles in the range $10^\circ <\theta_{\rm bar}
<25^\circ$, the 2:1 orbits naturally give rise to secondary peaks in
the line of sight velocity distributions at $v_{\rm los} \sim$ 200
km s$^{-1}$.  We have provided a pictorial explanation of this
phenomenon. 

Our interpretation is open to the objection that the method is not fully self-consistent. 
We have shown that the population of 2:1 orbits can generate the kinematic features to explain the data of \citet{2012Nidever}, but we have allowed the normalization of the density in these orbits to vary independently of the N-body model from which they were extracted. 
This though is unlikely to be a serious concern, as the range of self-consistent equilibria for bars is wide and solutions will exist using different relative populations for the orbital families that comprise the bar. 
This has been demonstrated explicitly for the related problem of static triaxial ellipsoids \citep[e.g.,][]{Statler1987,Hunter1995}. For bars, the ability to exchange orbits without changing the density is indicated by both the non-unique decomposition into the classical and regular components found by \citet{Hafner2000} and the range of made-to-measure models reproducing the density inferred from the VVV survey \citep{Wegg2013} found by \citet{Portail2015a}. As we have mentioned above, the other resonant orbit families also produce rich structure in their velocity distributions. If we combine the 2:1, 3:1 and 5:2 orbits in our distributions, the high velocity peaks (which are somewhat more obscured) are dominated by the 2:1 orbits and so follow the trends outlined above to suggest $\theta_{\rm bar}\approx15^{\circ}$.
Another important consideration is the range of distances probed by observations. Given the non-uniform distribution of dust towards the Galactic Bulge, it is certain that different fields are reaching different distances. Indeed the range of distances reached may vary significantly through just one field. The lower right panel of Figure \ref{fig:21BarOrbit} shows that the highest velocity 2:1 stars are placed at a distance close to $R_{0}$. There is also likely to be some variation in the relative fractions of the different orbit families as one varies the distance probed. It is clear then that distance has an important role in shaping the velocity distributions.

\subsection{Puzzles in the Data}
Some issues remain about the high velocity peaks. 
The first is why symmetry is not seen between positive and negative
latitude fields. Peaks are observed in the field at (4.3,-4.3) but not
in the field at (4.3,4.3), the same goes for the fields at (5.7,-2)
and (5.7,2).  The obvious explanation for such a difference is that
the distances being probed differs between north and south. 
It is known that extinction in the north is greater than in the south
\citep{Gonzalez2012}, which offers a possible explanation for the
disparity. However, according to APOGEE estimates, although the
average extinction is indeed greater for the field at (4.3,4.3) than
for the field at (4.3,-4.3), the opposite is the case for the fields
at (5.7,2) and (5.7,-2). The extinction data is at its
most uncertain close to the plane of the disc, so it is unclear if this is causing the difference between symmetric fields. Another point to consider is that in the
plane, where extinction is highest, strong peaks are seen. As we show below, a  likely explanation here is that the number density of bar supporting orbits
is higher in these regions. 

The second major concern is the differences between different Bulge
surveys. The BRAVA, ARGOS and GIBS surveys don't report the detection
of cold components in their $v_{\rm los}$ distributions. The presence of the peaks in the APOGEE
data should be resolutely confirmed on analysis of the
post--commissioning data. That the cold peaks are seen in one survey
and not the others provides a possible clue as to the nature of bar
supporting orbits. The clue lies in the different observing
strategies and selection functions for the data.
The color selections are similar, with each survey using cuts on
$J-K$ color, but the magnitude ranges are not the same.
These surveys each cover different footprints, with only
APOGEE and GIBS observing a significant number of fields below $\left|
b \right|=4^\circ$. Most of our sample of resonant 2:1 orbits are confined close to the plane with $\left| b \right|<5^\circ$ (corresponding to $\sim$0.75 kpc at $R_{0}$). Figure \ref{fig:vlos_b} (top left) shows the projected surface density (on a log scale; contours at 0.1 dex) for the 2:1 family, where we have averaged over 0.48 Gyr. In the following panels we show the $v_{\rm los}$ distributions for stripes in longitude ($-10^\circ<l<10^\circ$) centred on $b=0^\circ,-2^\circ$ \& $-4^\circ$ (A, B \& C respectively; black histograms indicate a strip of width $\triangle b=1^\circ$ and red, a strip of width $\triangle b=0.5^\circ$).
There is a stark difference between the distribution at $b=0^\circ$ (A) and $b=-4^\circ$ (C). The strong peaks, clearly visible in the mid-plane, rapidly decrease in prominence as latitude is increased. At $b=-4^\circ$ there is only a small hint of a high velocity component at $\sim200$ km/s. This, along with the low density of 2:1 orbits, makes the detection of a cold peak much less likely at higher latitudes. It seems reasonable then that the BRAVA and ARGOS surveys don't report the detection of cold peaks. This effect is unlikely to be the only important difference between the surveys. As we have shown above (Figure \ref{fig:21BarOrbit}), the distances probed will also have a significant impact on the shape of the $v_{\rm los}$ distributions. 

A more subtle difference is that the surveys observe different types of stars
(GIBS \& ARGOS: mainly red clump; APOGEE \& BRAVA: mainly
M-giants)\footnote{It has been subsequently shown by \citet{Aumer2015} that the APOGEE selection function is indeed selecting young stars.}. Since the targets that make up each survey differ in
spatial (survey footprint, distance), temporal (ages) and chemical
(color, magnitude) attributes, the question of why cold peaks are
seen in one survey and not others is certainly challenging (it should also be noted that the observing strategy may be selecting stars with a low velocity dispersion, thereby thinning the overall distributions to reveal the cold peak). A better
question might be whether APOGEE has stumbled on a selection strategy
that preferentially selects bar stars, and if so, can the strategy be
shown to be consistent with chemical and dynamical models of the
Galactic Bulge?

We have so far not discussed in detail the effect of fore-/background or chaotic stars on the $v_{\rm los}$ distributions. \citet{Aumer2015} suggest that the fore-/background stars play an important role in shaping the ``main" distribution. The method outlined in M15 currently has no way of reliably identifying chaotic orbits but, in any case, they are unlikely to generate strong, long-lived features. \citet{Debattista2015} have also recently suggested that substructure, similar to that seen in the APOGEE data, can be generated by a rotating nuclear disc. Their kpc-scale nuclear disc provides an alternative explanation for the observed kinematics. 

So far, we have considered the cold peak to be a real and statistically significant feature. \citet{Li2014} showed that peaks can be generated by under-sampling the wings of a broad $v_{\rm los}$ distributions. It is unlikely however that similar features would be randomly drawn in a number of independent fields. \citet{Debattista2015} have also recently shown that such features are unlikely to arise due to Poisson noise. 

\section{Conclusions}
We have set out a framework for understanding the substructure seen in the APOGEE data. As was suggested by \citet{2012Nidever}, we have shown that the motions of stars along periodic bar orbits naturally give rise to cold peaks in $v_{\rm los}$ distributions. In particular, the 2:1 family are able to simultaneously generate peaks close to the observed values across a number of fields. This is a strong argument for the 2:1 family being the main driver behind the location of the peaks. Although uncertain, the relative fraction for the different families suggested by this simulation \citep[and previous works, e.g.,][]{Sparke1987} points to the 2:1 family being the most populous in barred potentials. 

With this hypothesis in mind, we can constrain the viewing angle of the bar with the locations the peaks generated by the 2:1 orbits to be $10^{\circ} \lesssim \theta_{\rm bar} \lesssim 25^{\circ}$. Although there remain uncertainties attached to this estimate, we are confident that large viewing angles can be discounted. The effect of fundamental bar parameters on the locations of the peaks remains to be explored in any detail. Uncertainties associated with the stellar distances, the effects of extinction and the selection function of the APOGEE survey make a complete comparison with models difficult. We have also shown, in agreement with \citet{Aumer2015}, that the higher order resonances can make a significant contribution to the $v_{\rm los}$ distributions, especially at high latitudes ($|b| > 2^{\circ}$). Close to the plane, however, the 2:1 family appears to play the dominant role and sets the locations of any high velocity peaks. 

Finally, this work has demonstrated the power of the method introduced in \citet{Molloy2015} for extracting periodic orbits from N-body simulations in which the underlying gravitational potential is unsteady. By dissecting the bar into its constituent periodic families, we are able to study their characteristic kinematical and spatial signatures. Forthcoming wide field spectroscopic surveys of the Galactic centre are likely to discover further streams, kinematic features and substructure. We are confident that the techniques of the paper will have an important role to play. 

\acknowledgements The authors acknowledge financial support from the CAS One
Hundred Talent Fund and NSFC Grants 11173002, 11333003, 11322326 and
11073037.  This work was also supported by the following grants: the
Gaia Research for European Astronomy Training (GREAT-ITN) Marie Curie
network, funded through the European Union Seventh Framework Programme
(FP7/2007-2013) under grant agreement no 264895; the Strategic
Priority Research Program ``The Emergence of Cosmological Structures''
of the Chinese Academy of Sciences, Grant No. XDB09000000; and the
National Key Basic Research Program of China 2014CB845700. This work made use of the super-computing facilities at Shanghai Astronomical Observatory.

\bibliography{myrefs2.bib}

\end{document}